\documentclass[letterpaper]{article}

\usepackage{natbib,alifeconf}  
\usepackage{amsmath}
\usepackage{url,hyperref,cleveref}
\usepackage{booktabs}
\usepackage{graphicx}
\usepackage {xhfill}
\usepackage{tcolorbox}
\usepackage{tabularx}
\usepackage{subcaption} 

\usepackage[whole]{bxcjkjatype}

\title{Evolution of Social Norms in LLM Agents using Natural Language}

\author{
   Ilya Horiguchi,
   Takahide Yoshida \and
   Takashi Ikegami
   \mbox{}\\
   Department of General Systems Studies, Graduate School of Arts and Sciences, University of Tokyo, Japan \\
   Email: horiguchi@sacral.c.u-tokyo.ac.jp
} 

\begin{document}

\maketitle

\begin{abstract}
    Recent advancements in Large Language Models (LLMs) have spurred a surge of interest in leveraging these models for game-theoretical simulations, where LLMs act as individual agents engaging in social interactions. This study explores the potential for LLM agents to spontaneously generate and adhere to normative strategies through natural language discourse, building upon the foundational work of Axelrod's metanorm games. Our experiments demonstrate that through dialogue, LLM agents can form complex social norms, such as metanorms—norms enforcing the punishment of those who do not punish cheating—purely through natural language interaction. The results affirm the effectiveness of using LLM agents for simulating social interactions and understanding the emergence and evolution of complex strategies and norms through natural language. Future work may extend these findings by incorporating a wider range of scenarios and agent characteristics, aiming to uncover more nuanced mechanisms behind social norm formation.
\end{abstract}

\section{Introduction}

In recent years, following the emergence of Large Language Models (LLMs), simulation studies based on game theory treating LLMs as individual agents have been conducted worldwide. For example, research on repeated games has revealed that while LLMs demonstrate excellent performance in games prioritizing self-interest, they exhibit suboptimal behavior in games requiring coordination~\citep{RepeatedGamesLLM}. Additionally, the NegotiationArena study \citep{NegotiationArena} showed how LLM agents can conduct complex negotiations through flexible dialogue in negotiation settings and significantly improve negotiation outcomes by employing specific behavioral strategies. Furthermore, the ALYMPICS study by \cite{ALYMPICS} proposed a systematic simulation framework for game theory research using LLM agents, aiming to simulate strategic interactions. However, previous studies have been limited to having agents choose simple options like cooperation (C) or defection (D), not fully utilizing the advantage of natural language capabilities. Therefore, this study aims to analyze the process of diverse strategy emergence as LLMs engage in rich discussions using natural language.

We focused on Axelrod's concept of metanorms from the perspective of spontaneous strategy emergence. A metanorm is a norm that punishes those who do not adhere to a certain norm~\citep{EvolutionaryNorms}. For instance, if there is a norm to punish cheaters, the corresponding metanorm would be to punish those who choose not to punish cheaters. Axelrod conducted simulation studies on how norms spontaneously emerge and stabilize. This concept has led to research by \cite{MetanormGames} exploring how various network structures affect the effectiveness of metanorms as a mechanism to promote cooperation in society, and Axelrod's own consideration that metanorms are an additional mechanism necessary for norm stability~\citep{ComplexityCooperation}. \cite{EvolutionaryPersonalityTraits} modeled the evolution of personality traits in game-theoretic relationships, demonstrating the evolution of cooperative behavior by using LLMs to use linguistic descriptions related to personality traits as genes. This showed that LLM agents can develop strategies through the evolution of natural language descriptions. These studies suggest the possibility that LLM agents can autonomously acquire diverse strategies such as cooperation and norms by evolving negotiation and strategic decision-making through language.

We explored whether metanorms could emerge among LLM agents through discussions using natural language. This study provides new insights into understanding the behavior of multi-agent LLMs. Particularly in the field of AI alignment, it is important for understanding how agents can spontaneously create autonomous norms in the presence of other values and social norms.

\section{Experiments}

\cite{Toolformer} demonstrated a method for LLMs to self-educate in the ability to use external tools. This capability enables autonomous use of various tools such as calculators, search engines, and calendars. In our experiments, as shown in \cite{NegotiationArena}, we adopted an approach that uses tags instead of tools, allowing LLM agents to operate as if inputting commands within the game. This enabled us to implement a simulation of the Norms Game using LLM agents. Specifically, we developed a tag system to instruct LLM agents' actions within the Norms Game. This system allows agents to select appropriate commands according to the game context and interactively incorporate information. We constructed an environment that enables the development of more complex strategies and the spontaneous emergence of metanorms.

\begin{figure}[ht]
\includegraphics[scale=0.2]{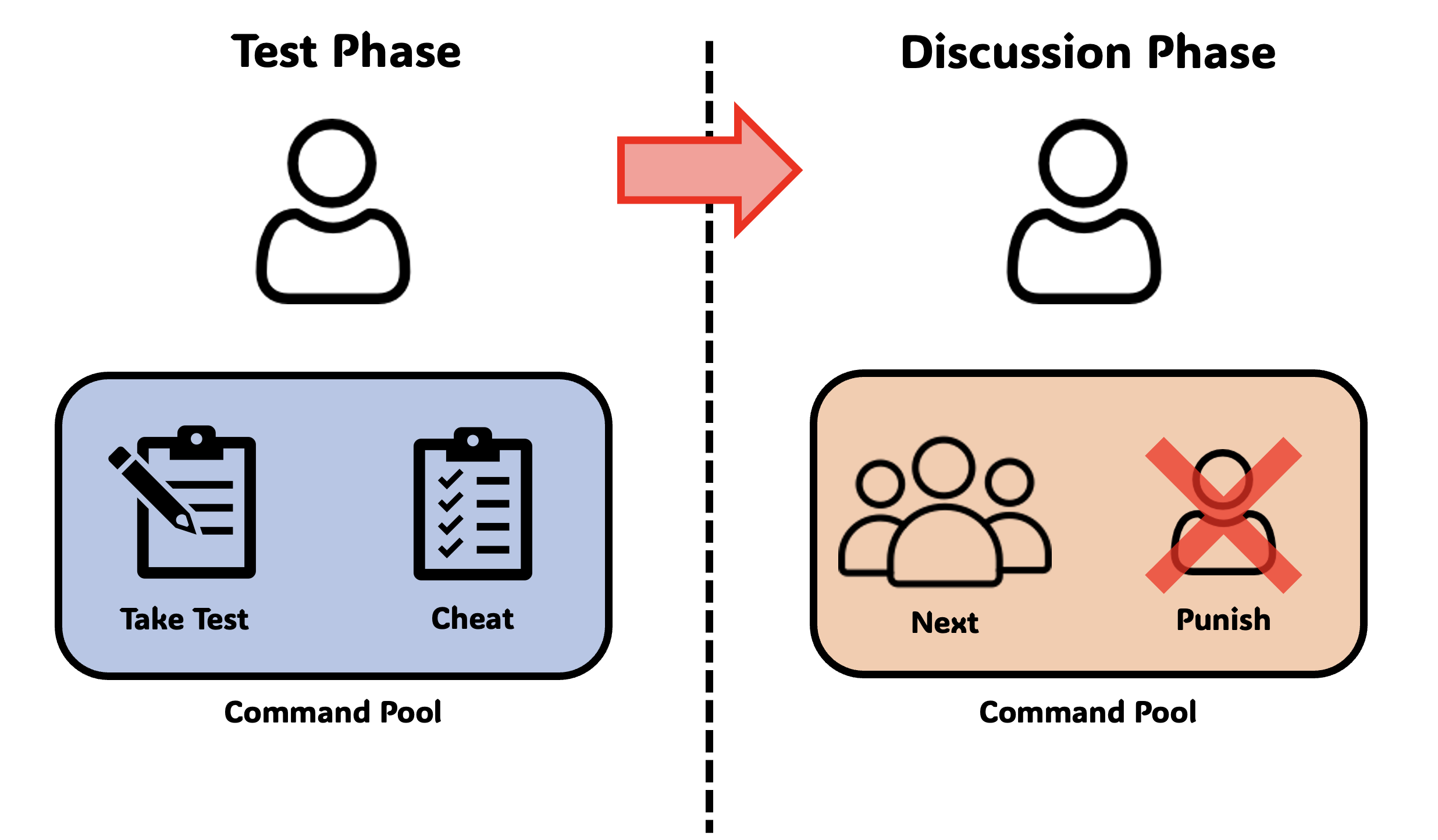}
\caption{Agent commands in a norms game}
\label{fig:command}
\end{figure}

\subsection{Norms Game}
In the Norms Game experiment, we modified Axelrod's experimental setup to allow agents to develop strategies more flexibly. Each agent can communicate their `vengefulness' and `boldness' levels in natural language, such as ``5 out of 7'', and then proceed with the game while embodying that personality.

All agents experience a test phase and a discussion phase. In the test phase, players choose between executing a test command or a cheat command (Figure \ref{fig:command}). When using the test command, a random value with a mean of 50 and variance of 10 is obtained as a score. In the case of cheating, players can gain an average of +30 points compared to taking the test. However, cheating is revealed when scores are announced to all agents.

In the discussion phase, agents take turns speaking, discussing each other's scores. In this phase, they can choose either the `next' command to designate who speaks next by name, or the `punish' command to penalize someone by name. The agent using the command pays 20 points to punish the target agent by -90 points. After a punish command, the next speaker is randomly determined. These values and settings conform to Axelrod's original paper. Table \ref{tab:score} shows an example of scores after command selection and discussion for each event occurrence.

\begin{table}[h]
\centering
\caption{Examples of gains for each command}
\label{tab:score}
\begin{tabularx}{\columnwidth}{Xccc}
\hline
Event       & Payoff per Event & Number of Events & Payoff \\ \hline
Baseline    & \( 50 \)               & 1                & \( 50 \)     \\
Cheat       & \(  +30 \)              & 1                & \(  +30 \)    \\
Punished    & \(  -90 \)              & 1                & \(  -90 \)    \\
PunishCost & \(  -20 \)              & 2                & \(  -40 \)    \\ \hline
\multicolumn{3}{r}{Score}      & \( -50 \)              \\ \hline
\end{tabularx}
\end{table}

\subsection{Evolution}
While referencing the experimental setup in Axelrod's original paper, we made several modifications in this study. In Axelrod's research, after all payoffs were calculated, agents with average payoffs in each epoch left one offspring, genes with payoffs one standard deviation above the mean doubled in number, and genes with payoffs one standard deviation below the mean were eliminated. Additionally, the strategy (boldness or vengefulness) of one individual was probabilistically rewritten. In the original paper, the population was fixed at 20, but in our study, we needed to reduce the number of agents due to the discussion phase where individuals speak one at a time. Therefore, we fixed the number of agents at 7 and adopted a method where the 3 agents with median scores when ranked by payoff were directly inherited, while the top 2 individuals' genes were doubled. Furthermore, one individual was randomly selected, and either their boldness or vengefulness was changed to a random value. This allowed us to observe more dynamic genetic changes and strategy evolution.

\section{Results}
This study used OpenAI's GPT-4\footnote{gpt-4-0125-preview, temperature=0.0}. The raw data of the results and the code used can be accessed on github\footnote{\url{https://github.com/NeoGendaijin/NormGame_LLM}}.
\subsection{Results of Controlling Vengefulness and Boldness}
In Axelrod's Norms Game, the following prompts corresponding to vengefulness and boldness were given to the agents.
\newtcolorbox{mybox}{colback=lime!30!white, colframe=lime}
\begin{mybox}
Character Traits:

        - Vengefulness: 5 out of 7
        
        - Boldness: 6 out of 7

Trait Descriptions:

- **Vengefulness** (omited)

- **Boldness** (omited)
\end{mybox}
We observed how discussions and Punishment progressed in groups with high/low vengefulness and high/low boldness. When the value was low, numbers 1-3 were uniformly randomly assigned, and when high, numbers 5-7 were uniformly randomly assigned. In each group, the discussion phase was set to have a maximum of 21 turn-takings, and the entire process was repeated 10 times. The following are the results.

\begin{figure}[ht]
\includegraphics[scale=0.40]{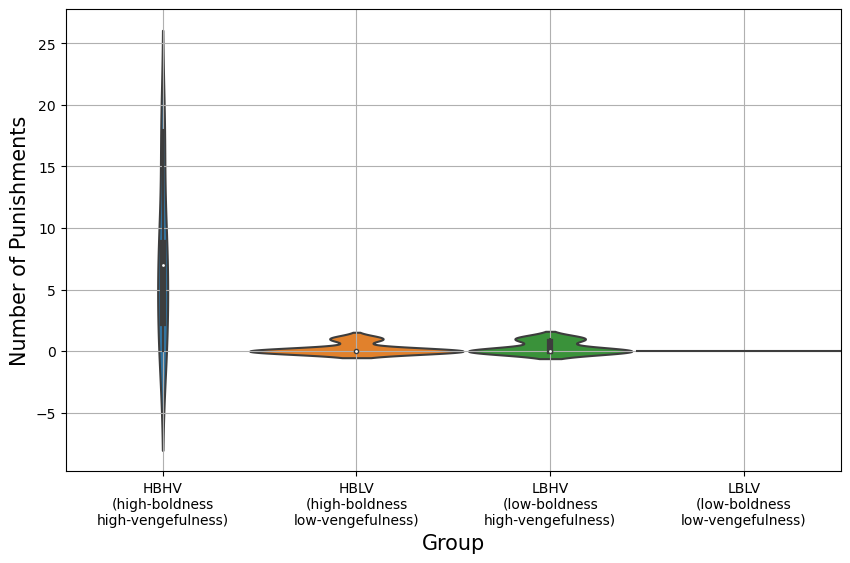}
\caption{Number of `punish' commands within discussions for each group}
\label{fig:violin}
\end{figure}

\begin{figure}[ht]
\includegraphics[scale=0.40]{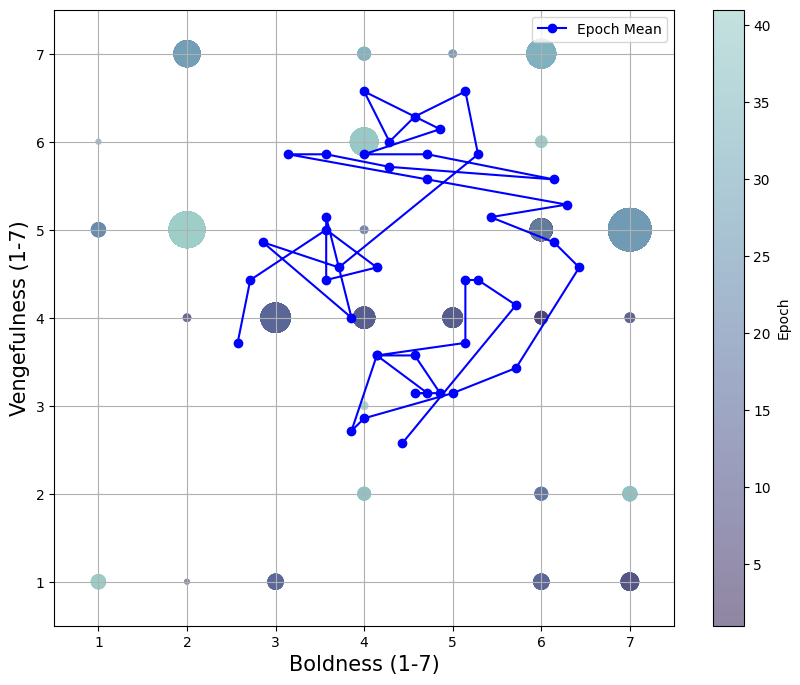}
\caption{Progression of vengefulness and boldness by epoch when introducing natural selection based on payoffs}
\label{fig:evo}
\end{figure}

\begin{figure*}[ht]
\centering
\includegraphics[scale=0.45]{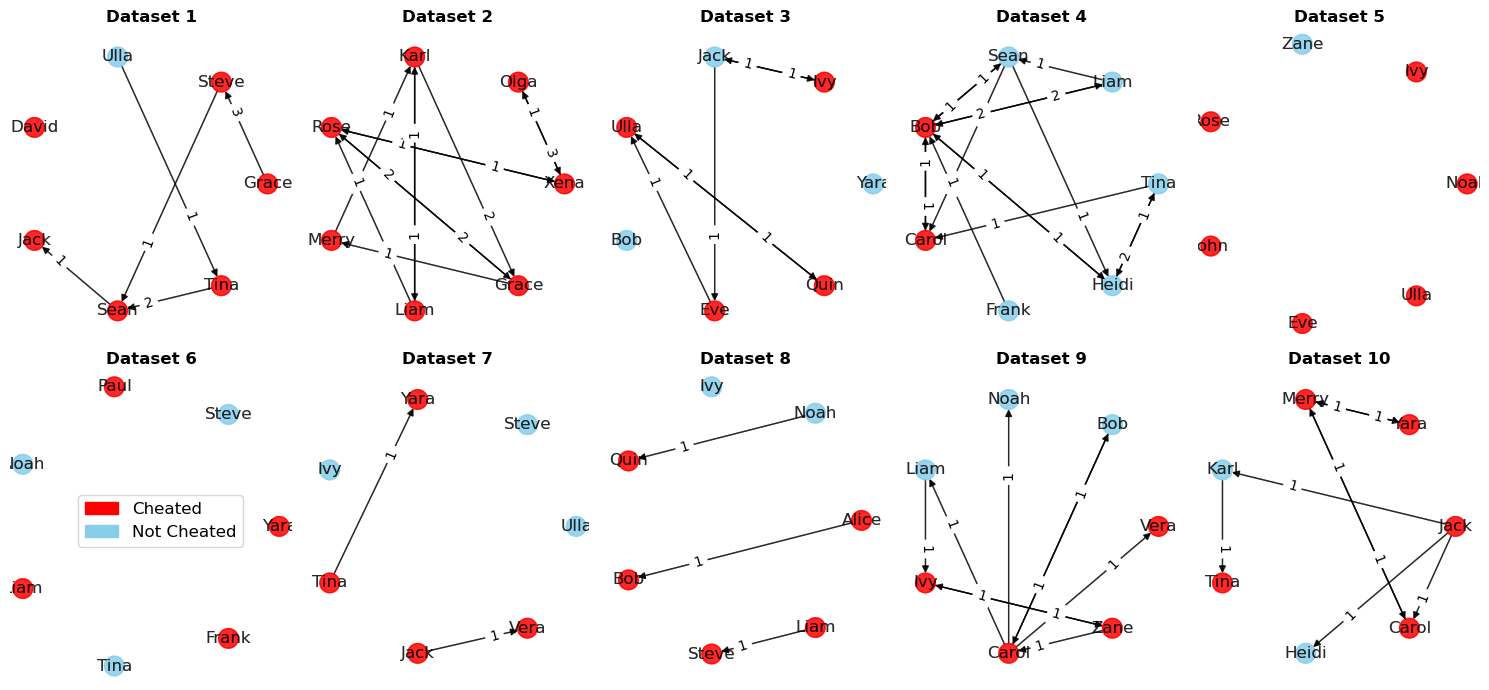}
\caption{Visualization of punishment network in high-vengefulness, high-boldness}
\label{fig:punish_net}
\end{figure*}

The distribution of the number of times Punishment was chosen in the discussion is shown in Figure \ref{fig:violin}. This figure shows that the Punish command was used most frequently in groups with high vengefulness and high boldness. The vertically elongated violin plot indicates large variations even within the same group. Conversely, in groups with low Vengefulness or Boldness, the number of Punishments was consistently low. To analyze the high vengefulness and high boldness group in more detail, Figure \ref{fig:punish_net} shows a visualization of the Punishment network with each agent as a node. Agents who cheated are colored red, and those who didn't are colored light blue. While there were rounds like Dataset2 where everyone punished each other, there were also rounds like Dataset5 and 6 where agents kept each other in check and no Punish commands were used at all. In Dataset4 and 9, when an agent who cheated was attacked by an agent who didn't cheat, another agent who had cheated retaliated with Punishment. Reading the agents' natural language judgments, it was observed that they adopted a strategy of protecting agents who used the same cheating strategy to increase their own scores. Similarly, in Dataset1 and 2, where almost everyone cheated, agents who punished cheating were themselves punished. This was because punishing when everyone cheated was judged by other agents as behavior that disrupted order. This principle of action is a type of meta-norm as described by Axelrod, and it can be considered to have naturally emerged from group discussions using natural language.
\subsection{Results of Evolution Based on Payoffs}
Figure \ref{fig:evo} shows the results of an experiment on natural selection based on payoffs for a group with random vengefulness and boldness. In each epoch, the discussion phase was set to have a maximum of 7 turn-takings (an average of 1 per person), and 40 epochs were executed. Circles of sizes corresponding to the population of each epoch are plotted on the grid. It can be seen that the initially scattered distribution gradually converges. Also, the average of 7 agents can be seen circulating around intermediate values for both vengefulness and boldness. The number of punishments evolved to increase once and then settle down. When cheating occurred, it was punished by agents with moderate vengefulness and was eliminated.

\subsection{Natural Language}
\begin{figure}[ht]
\includegraphics[scale=0.27]{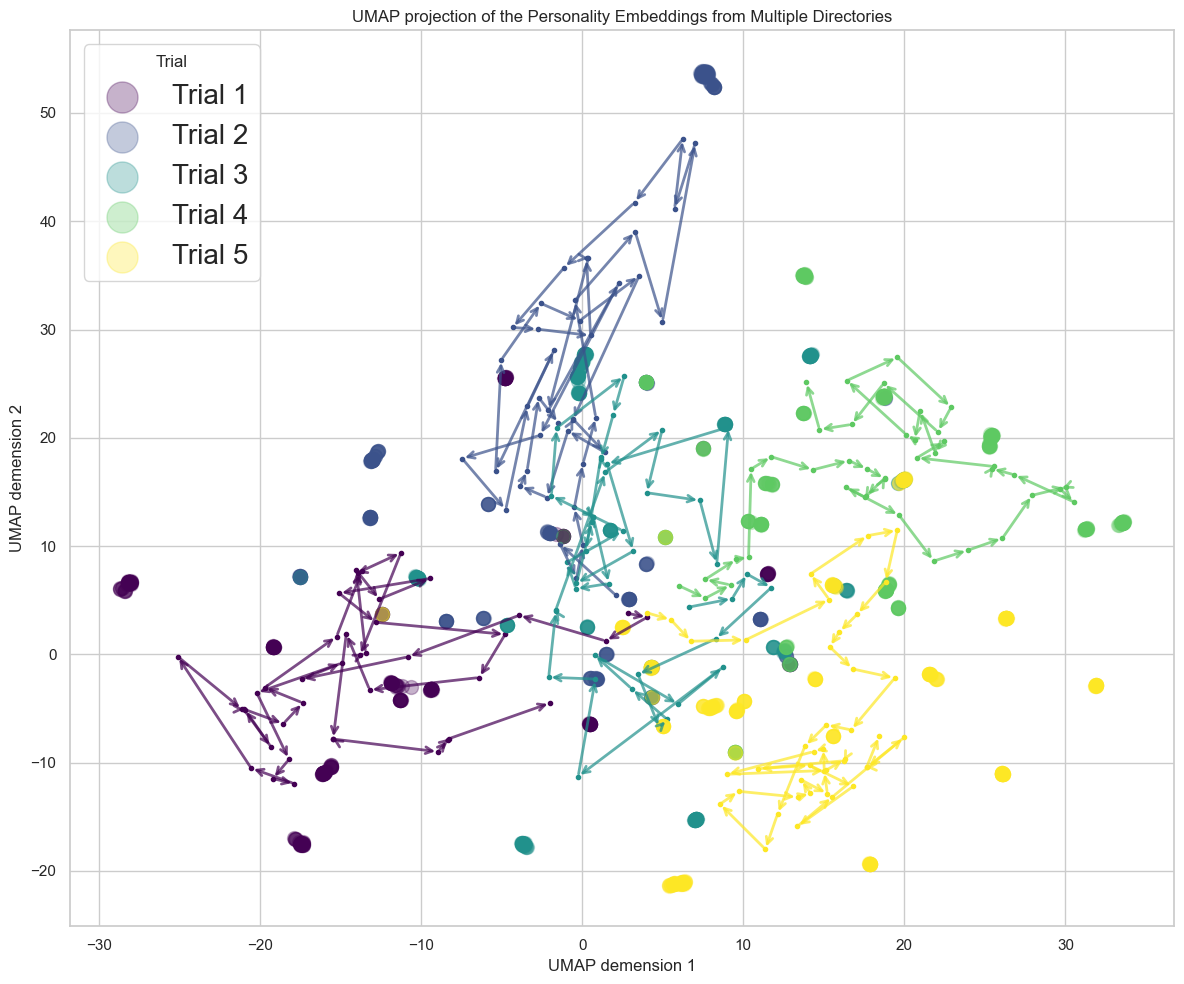}
\caption{Embedding plot across all five trials. Arrows indicate the progression of means for each epoch.}
\label{fig:evo_embedding}
\end{figure}

\begin{figure}[ht]
\includegraphics[scale=0.28]{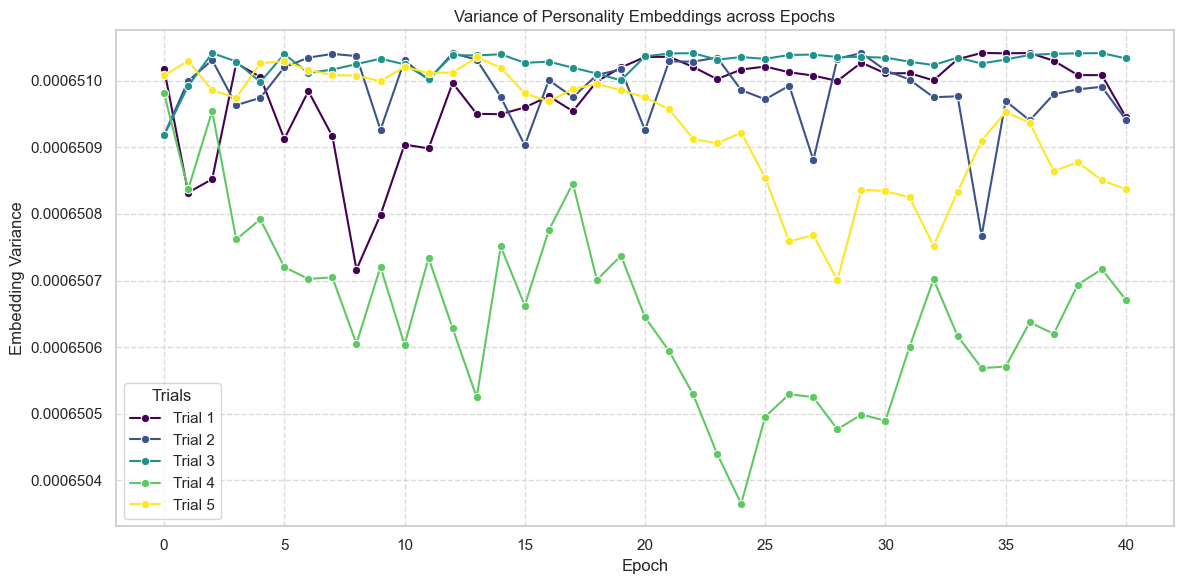}
\caption{Progression of embedding variance over 40 epochs.}
\label{fig:evo_sd}
\end{figure}

Attempts to incorporate evolutionary algorithms into LLMs have been explored in numerous studies \citep{Guo2023-ev}. In previous research \citep{EvolutionaryPersonalityTraits}, the personalities of LLMs were evolved using genetic algorithms. This approach simulated natural language variations by having LLMs rephrase expressions. In our experimental setup, we departed from Axelrod's original parameters and introduced expressions not constrained by vengefulness and boldness, evolving personalities and strategies through rephrasing. Each personality was expressed in approximately 10 words, with the rephrasing also performed by the LLM.
To initiate each epoch, we had the LLM randomly generate personalities and select from this personality pool. Within each epoch, 21 rounds were executed as in previous experiments. Based on the final scores, personalities were inherited through a natural selection process for the next generation. We examined the evolutionary trajectory over 40 epochs, with each epoch consisting of 21 rounds.
Figure \ref{fig:evo_embedding} shows the projection of average personality embeddings across five trials using UMAP. While trials 1, 2, and 3 appear to converge to similar points on the UMAP, trials 3 and 4 transition to different regions. Figure \ref{fig:evo_sd} illustrates the variance of embeddings, confirming that all members do not converge to identical personalities but evolve with a certain spread in the embedding space.

\begin{figure}[ht]
\includegraphics[scale=0.30]{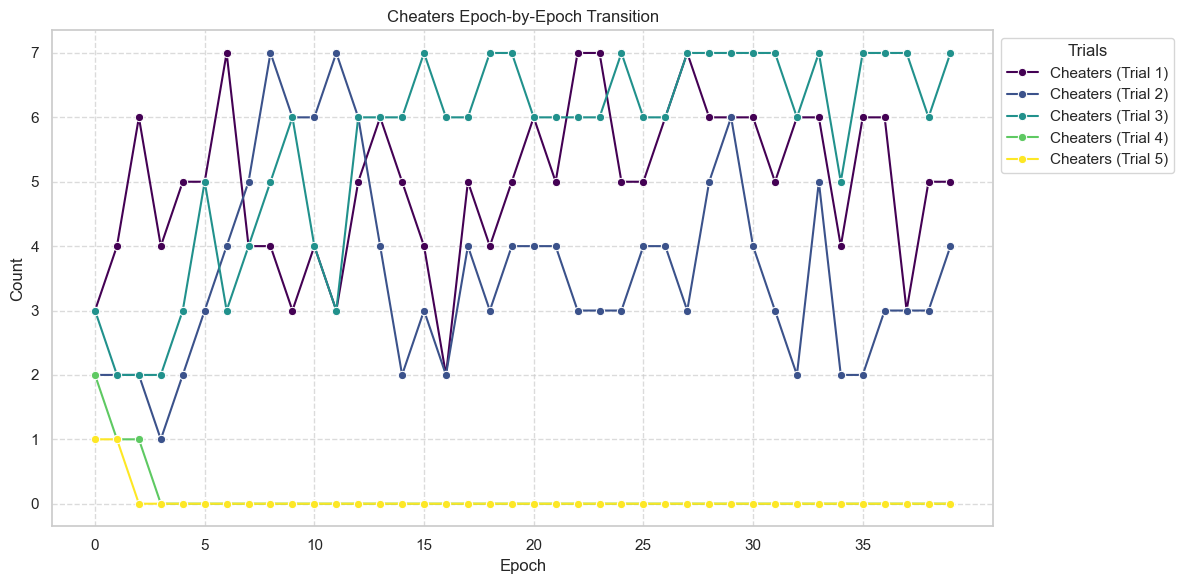}
\caption{Progression of members' cheating behavior over 40 epochs.}
\label{fig:evo_chaet}
\end{figure}

\begin{figure}[ht]
\includegraphics[scale=0.30]{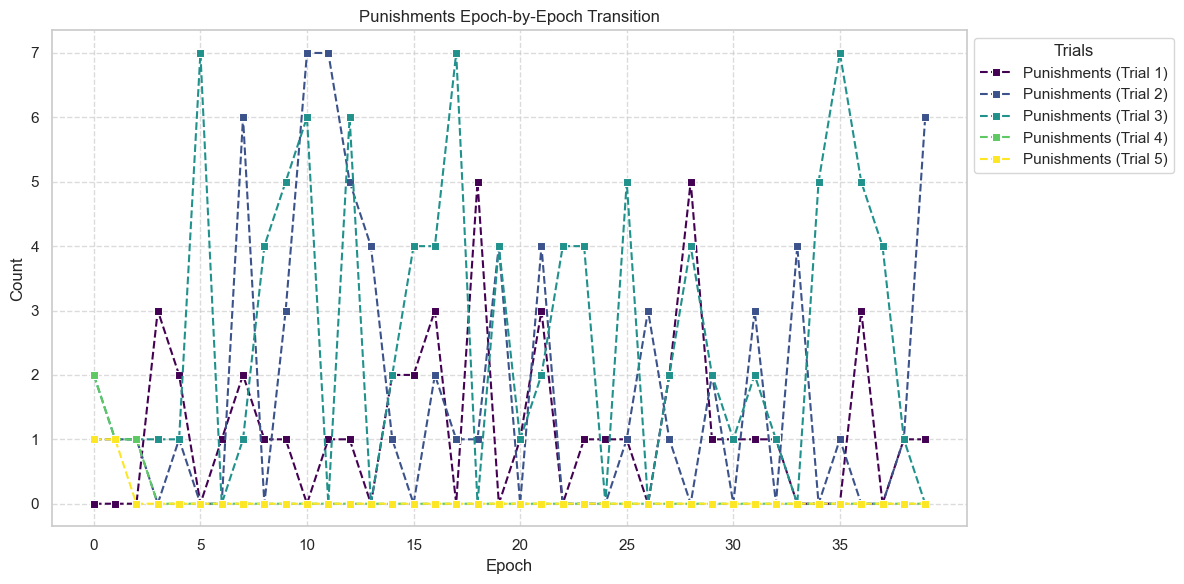}
\caption{Progression of members' punishment behavior over 40 epochs.}
\label{fig:evo_punish}
\end{figure}

Focusing on behavior, Figure \ref{fig:evo_chaet} reveals that trial 3 has a high proportion of members engaging in cheating. In contrast, trials 4 and 5 show that cheaters are quickly eliminated, forming communities that do not cheat. Figure \ref{fig:evo_punish} demonstrates that trials and epochs with higher rates of cheating generally exhibit higher rates of punishment.
The evolutionary outcomes depend on the initial personality pool, and even starting from similar personality pools, we observe the emergence of diverse strategies. This approach to evolving natural language expressions provides rich insights into the development of complex social norms and strategies, albeit with challenges in analysis and interpretation that future research will need to address.

\section{Conclusion}

This study explored the emergence of normative strategies through rich natural language discussions in game theory-based simulations using LLM agents. Based on Axelrod's meta-norm game, we implemented a Norms Game where agents engage in discussions, revealing the formation of complex social norms such as punishment for cheating and meta-norms for not punishing.Our experiments demonstrated that agents' personality traits, particularly vengefulness and boldness, significantly influence behavior choices and punishment patterns. The evolution experiment based on payoffs suggested that balancing these traits is crucial for agents' survival strategies, with the group being dominated by agents who neither cheated excessively nor punished others too frequently. The use of natural language in evolving strategies introduced both richness and challenges to our study. While it allowed for more nuanced and complex strategy development, it also presented difficulties in analysis and interpretation. Future research should address these challenges, potentially by exploring the effects of linguistic ambiguity on strategy emergence, possibly leading to interesting and unexpected outcomes. Incorporating elements of false belief tasks could be valuable to examine how strategies evolve in situations with incomplete information, and how agents predict and respond to others' mental states. It would also be interesting to investigate how the same personality traits (e.g., vengefulness and boldness) manifest in different game scenarios, providing insights into the transferability of evolved strategies. Analyzing how group size affects strategy evolution, particularly examining the differences between smaller groups and larger ones where subgroup formation might occur, could yield valuable insights. Additionally, applying psychological frameworks like the Big Five personality model to track how agent personalities evolve across different game scenarios might yield insights of interest to psychologists.

These future directions could provide a more comprehensive understanding of how complex social norms and strategies emerge and evolve through natural language interactions. By addressing the challenges of analysis and interpretation in language-based evolutionary models, we can further refine our approach to studying the dynamics of social norm formation in AI systems.This research confirms that social interaction simulations using LLM agents are an effective method for understanding the emergence and evolution of complex strategies and norms through natural language. As we continue to refine these methods, we expect to gain deeper insights into the mechanisms underlying diverse social norms, potentially informing both AI development and our understanding of human social dynamics.


\footnotesize

\bibliographystyle{apalike}
\bibliography{citation} 

\end{document}